\documentclass[12pt]{article}

\catcode`\@=11
\@addtoreset{equation}{section}

\global\arraycolsep=2pt
\usepackage{mathrsfs,amsbsy,amssymb,latexsym,amsfonts,amsmath,cite}
\usepackage{graphicx,color}

\allowdisplaybreaks

\newcommand{\bea}{\begin{eqnarray}}
\newcommand{\eea}{\end{eqnarray}}
\newcommand{\beq}{\begin{equation}}
\newcommand{\eeq}{\end{equation}}

\def\eqref#1{Eq.~(\ref{#1})}

\def\Eq#1{\begin{equation} #1 \end{equation}}
\def\Eqr#1{\begin{eqnarray} #1 \end{eqnarray}}
\def\Eqrsubl#1#2{\begin{subequations}\label{#1}\Eqr{#2}\end{subequations}}

\newcommand{\nn}{\nonumber}
\newcommand{\pd}{\partial}
\newcommand{\vect}[1]{\!\!\!\mbox{ \boldmath $#1$}}

\def\Xsp{{\rm X}}
\def\Ysp{{\rm Y}}
\def\Zsp{{\rm Z}}
\def\X5sp{{\rm X}_5}
\def\Y3sp{{\rm Y}_3}
\def\Z3sp{{\rm Z}_3}
\def\Msp{{\rm M}}

\def\lap{{\triangle}}
\def\e{{\rm e}}

\begin{document}

\begin{flushright}
\parbox{4cm}
{KUNS-2455}
\end{flushright}

\vspace*{0.5cm}

\begin{center}
{\Large \bf Dynamical F-strings intersecting D2-branes \\ in type IIA supergravity}
\vspace*{1.5cm}\\
{\large Kunihito Uzawa$^{\ast}$
and Kentaroh Yoshida$^{\dagger}$
} 
\end{center}
\vspace*{0.25cm}
\begin{center}
$^{\ast}${\it Department of Physics,
School of Science and Technology, \\ 
Kwansei Gakuin University, Sanda, Hyogo 669-1337, Japan
}%
\vspace*{0.5cm} \\ 
$^{\dagger}${\it Department of Physics, Kyoto University \\ 
Kyoto 606-8502, Japan
} 
\end{center}
\vspace{1cm}
\begin{abstract}
We present time-dependent exact solutions composed 
of F-strings intersecting D2-branes in type IIA supergravity. 
The F-strings tend to approach each other with time 
in the six-dimensional space transverse to both F-strings and D2-branes. 
In general, a singularity appears before collision. 
An exceptional case is that the charges are the same and five directions 
in the transverse space are smeared out. 
Then we argue some applications of the solutions 
in building cosmological models. 
The possible models are classified based on compactifications of the 
internal space. 
All of them give rise to the Friedmann-Robertson-Walker universe 
with a power-law expansion. 
\end{abstract}

\setcounter{footnote}{0}
\setcounter{page}{0}
\thispagestyle{empty}

\newpage

\section{Introduction}

A fascinating issue in general relativity and string theory is to study 
time-dependent brane solutions. The time evolution of them is intimately 
related to the D-brane dynamics in string theory and also there are some 
potential applications to cosmology and black hole physics. 
We are concerned here with colliding brane solutions. 
The original colliding brane solutions have been found by Gibbons 
{\it et al}. \cite{GLP}. 
For some generalizations of the solutions, see 
\cite{GLP, CCGLP, KU1, KU2, BSU, Maeda:2009tq,
 MOU, Gibbons:2009dr, Maeda:2009ds, Maeda:2010ja, Minamitsuji:2010fp, 
 Maeda:2010aj, Minamitsuji:2010kb, Nozawa:2010zg, Minamitsuji:2010uz, 
 Maeda:2011sh, Minamitsuji:2011jt, Maeda:2012xb, Minamitsuji:2012if}\,.  
A remarkable point is that the solutions are time-dependent but still exact 
and analytic. 

\medskip 

The colliding brane solutions are constructed basically by generalizing 
static brane solutions in supergravities
\cite{GLP, CCGLP, KU1, KU2, BSU, MOU, Minamitsuji:2010fp, 
 Minamitsuji:2010kb, Nozawa:2010zg, Minamitsuji:2010uz, 
 Maeda:2011sh, Minamitsuji:2011jt, Maeda:2012xb, Minamitsuji:2012if}. 
As a general property, the solutions give rise to Friedmann-Robertson-Walker 
(FRW) universes typically by regarding a homogeneous and isotropic part of 
the solution 
as our four-dimensional spacetime. 
Also, some colliding solutions may provide black holes 
in a FRW universe when three of the transverse directions to the D-branes 
are regarded as the three spatial directions of our world.  

\medskip 

Although the colliding brane solutions have interesting aspects, 
there are some problems. 
As a feature of the colliding solutions, for example,
a warp factor in the metric is a linear function of time for 
a nontrivial dilaton\footnote{  
When the dilaton is constant or there exists no dilaton in the theory as 
in the eleven-dimensional supergravity, 
the time-dependence is not restricted to be linear and higher-order 
dependence is possible.}.  
Hence, even in the case of the fastest expansion, the power is too small 
to exhibit a realistic expansion law as in the matter dominated era or 
in the radiation dominated era. 
A possible resolution is to include some additional matter fields. 

\medskip 

As another application of time-dependent solutions, one may consider 
condensed matter systems. An example is a Lifshitz point \cite{LP}. 
The static solutions dual to a Lifshitz point 
have originally been shown in \cite{Kachru}. 
The embedding of them into string theory has been carried out in 
\cite{BN,DG}. There exist D3-brane solutions that 
give rise to the Lifshitz space as a near-horizon geometry \cite{CH}. 
In the previous work \cite{Uzawa:2013koa}, we have considered a 
time-dependent generalization of the solution found in \cite{CH}.  
Then dynamical Lifshitz-type solutions \cite{Uzawa:2013koa} have been 
constructed with a compactification \cite{DG}.  
The solutions describe the time evolution from the Lifshitz space to 
an AdS space and it may be interpreted as an aging phenomenon in 
condensed matter systems. 

\medskip 

In this paper, we first generalize the D3-brane solutions constructed 
in \cite{Uzawa:2013koa} to {\it multicenter} D3-brane solutions with 
{\it multiple} waves. Then we consider a T-dual picture of the generalized 
D3-brane solutions (without the compactification to the Lifshitz space). 
With a slight extension, we obtain a brane system composed of dynamical 
F-strings intersecting D2-branes. In the six-dimensional space transverse 
to both F-strings and D2-branes (called the overall transverse space),   
the positions of the D2-branes are fixed at some points while F-strings 
are dynamical and tend to approach each other with time. In general, 
a singularity appears before collision. An exceptional case is that the 
F-string charges are the same and five directions of 
the overall transverse space are smeared out. 
We also discuss some cosmological aspects of the solutions 
towards building a realistic cosmological model. 
The possible models are classified based on compactifications of the 
internal space. All of them give rise to the FRW universe with a power-law 
expansion. 

\medskip 

This manuscript is organized as follows. 
In Sec.~\ref{wave}, we first present time-dependent multicenter 
D3-brane solutions with multiple waves in type IIB supergravity. 
In Sec.~\ref{D2}, by performing a T-duality, 
we construct dynamical solutions composed of F-strings intersecting 
D2-branes in type IIA supergravity. The solutions describe collision of 
F-strings when the F-string charges are the same and five directions of 
the six-dimensional transverse space are smeared out. 
In Sec.~\ref{cos} we argue cosmological implications of the solutions. 
We also give a classification of lower-dimensional theories with 
compactifications. 
Section \ref{Conclusion} is devoted to conclusion and discussion.

\section{Time-dependent D3-brane solutions with waves} 
\label{wave}

Let us present time-dependent D3-brane solutions with multicenters  
and multiple waves in type IIB supergravity. 

\medskip 

We restrict ourselves to the solutions that contain 
the dilaton $\phi$, the axion $\chi$ 
and the self-dual five-form field strength $F_{(5)}$\,. 
Then the relevant field equations are given by
\Eqrsubl{D3p:equations:Eq}{
&&\hspace{-1cm}R_{MN}=\frac{1}{2}\pd_M\phi\pd_N\phi
+\frac{1}{2}\e^{2\phi}F_{M} {F_N}
+\frac{1}{4\cdot 4!}F_{MA_2\cdots A_5} 
{F_N}^{A_2\cdots A_5}\,,
   \label{D3p:Einstein:Eq}\\
&&\hspace{-1cm} F_{(1)} \equiv d\chi\,,~~~~~
d\left[\e^{2\phi}\ast F_{(1)}\right]=0\,,~~~~~
dF_{(5)}=0\,,~~~~~F_{(5)}=\ast F_{(5)}\,.
   \label{D3p:gauge:Eq}
}
Here $\ast$ is the Hodge operator in ten dimensions. 

\medskip 

Now let us suppose the following ansatz: 
\Eqrsubl{D3p:ansatz:Eq}{
ds^2&=&h^{-1/2}(t, x, y^i, z^a)\left[-\left\{
2-h_{\rm W}(t, x, y^i, z^a)\right\}dt^2
+2\left\{1-h_{\rm W}(t, x, y^i, z^a)\right\} dtdx
\right. 
\nn\\
&&
\left. 
+h_{\rm W}(t, x, y^i, z^a)\,dx^2+\gamma_{ij}(\Ysp)\,dy^idy^j
\right]
+h^{1/2}(t, x, y^i, z^a)\,u_{ab}(\Zsp)\,dz^adz^b\,,
 \label{D3p:metric:Eq}\\
\phi&=&\phi_0\,,
  \label{D3p:phi:Eq}\\
F_{\left(1\right)}&=&-\frac{k}{\sqrt{2}}\left(dt-dx\right)\,,
  \label{D3p:F1:Eq}\\
F_{\left(5\right)}&=&(1\pm\ast) d\left[h^{-1}(t, x, y^i, z^a)\wedge dt
\wedge dx\wedge\Omega(\Ysp)\right]\,. 
  \label{D3p:F5:Eq}
}
The two-dimensional metric $\gamma_{ij}(\Ysp)$ and the six-dimensional metric $u_{ab}(\Zsp)$ 
depend only on $y^i~(i=1,2)$ and $z^a~(a=1,\ldots,6)$\,, respectively. 
Then $\phi_0$ and $k$ are constant. 
The volume two-form  $\Omega(\Ysp)$ is given by 
\Eqr{
\Omega(\Ysp)=\sqrt{\gamma}\,dy^1\wedge dy^2\,, \qquad \gamma \equiv \det(\gamma_{ij})\,.
   \label{D3p:volume:Eq}
}
The constant $k$ is interpreted as the momentum of wave on 
the D3-branes in our later discussion.

\medskip 

The above ansatz reduces the field equations to the following form:  
\Eqrsubl{D3p:fields2:Eq}{
&&R_{ij}(\Ysp)=0\,,~~~~R_{ab}(\Zsp)=0\,,
   \label{D3p:Ricci2:Eq}\\
&&h=h(z^a)\,,~~~~h_{\rm W}=w_0(t)+w_1(x)+w_2(y^i, z^a)\,,
   \label{D3p:h2:Eq}\\
&&\lap_{\Zsp}h=0\,,~~~\pd_t^2w_0=0\,,~~~\pd_x^2w_1=0\,,
~~~\triangle_{\Zsp}w_2+h\left(\lap_{\Ysp}w_2
   +\frac{k^2}{2}\,\e^{2\phi_0}\right)=0\,,
   \label{D3p:warp-w:Eq}
 }
where $\lap_{\Ysp}$ and $\lap_{\Zsp}$ denote the Laplace operators on Y and Z 
spaces, respectively. 

\medskip 

From now on, we shall focus upon a special case by imposing the conditions, 
\Eq{
\gamma_{ij}=\delta_{ij}\,,~~~u_{ab}=\delta_{ab}\,,
~~~\lap_{\Ysp}w_2=0\,,~~~k\ne 0\,,
 \label{D3p:flat:Eq}
 }
where $\delta_{ij}$, $\delta_{ab}$ are the two-, six-dimensional 
Euclidean metric, respectively. 

\medskip 

Then, by solving the field equations, $h_{\rm W}$ and $h$ are given by  
\Eqrsubl{D3p:solutions1:Eq}{
\hspace{-0.9cm}h_{\rm W}(t, x, y^i, z^a)&=& a_0 + a_1\,t + a_2\,x
+b_i\,y^i+\sum_{\alpha=1}^{N}\frac{M_{\alpha}}
{2|\,\vect{z}-\,\vect{z}_{\alpha}|^4}\nn\\
&&+\frac{k^2}{8}\e^{2\phi_0}\sum_{\ell=1}^{N'}
\left(\frac{L_\ell}{|\,\vect{z}-\,\vect{z}_\ell|^2}
-\frac{|\,\vect{z}-\,\vect{z}_\ell|^2}{3}\right)\!,~~~~
M_{\alpha}=\frac{(2\pi)^3g_s^2\,N_\alpha\,l_s^8}{R_x\,V_{\Ysp}}\,,
 \label{D3p:solution-w:Eq}\\
\hspace{-0.9cm}
h(z^a)&=&1+\sum_{\ell=1}^{N'}\frac{L_\ell}{|\,\vect{z}-\,\vect{z}_\ell|^4}\,,
~~~~~L_\ell=2^2\,\pi\,\Gamma(2)\,g_s\,N_\ell\,l_s^4\,,
 \label{D3p:solution-h:Eq}
}
where $a_0$, $a_1$, $a_2$, $b_i$, $N_\alpha$ and $N_\ell$ 
are constant parameters; $g_s$ is 
the asymptotic string coupling constant; $l_s$ is the string length, 
related to the string tension $(2\pi\alpha')^{-1}$ as $\alpha'=l_s^2$\,; 
$R_x$ is the radius of $x$\,; and $V_{\Ysp}$ is the volume of Y. 
Here $M_{\alpha}~ (\alpha=1\,,\cdots\,, N)$ are masses of traveling waves 
located at  $\vect{z}_\alpha$ and $L_{\ell}~ (\ell=1\,,\cdots\,, N')$
are masses of D3-branes located at $\vect{z}_\ell$ and the $x$ direction 
is compact with period $2\pi R_x$\,. 

The distances in the six-dimensional space with $z^1,\ldots, 
z^6$ are defined as 
\Eqrsubl{D3p:z:Eq}{
&&|\,\vect{z}-\,\vect{z}_\alpha| 
\equiv \sqrt{\left(z^1-z^1_\alpha\right)^2+\left(z^2-z^2_\alpha\right)^2
+\cdots+
\left(z^6-z^6_\alpha\right)^2}\,,\\
&&|\,\vect{z}-\,\vect{z}_\ell|
\equiv \sqrt{\left(z^1-z^1_\ell\right)^2+\left(z^2-z^2_\ell\right)^2+\cdots+
\left(z^6-z^6_\ell\right)^2}\,.
}
The solutions with (\ref{D3p:solution-w:Eq}) and (\ref{D3p:solution-h:Eq}) 
describe 
$N$-center D3-branes with $N'$ waves. The directions along which the 
D3-branes and the waves extend are listed in Table \ref{D3p}. 
The time-dependent D3-brane solutions 
found in \cite{Uzawa:2013koa} correspond to the case with $N=N'=1$\,. 

\begin{table}[h]
\vspace*{0.5cm}
{\scriptsize
\begin{center}
\begin{tabular}{|c|c|c|c|c|c|c|c|c|c|c|}
\hline
Branes&0&1&2&3&4&5&6&7&8&9\\
\hline
D3 & $\circ$ & $\circ$ & $\circ$ & $\circ$ &  &  &
&  &  &  \\
\cline{2-11}
Wave & $\circ$ & $\star$ & & & & & & &  &  \\ 
\cline{2-11}
$x^N$ & $t$ & $x$ & $y^1$ & $y^2$ & $z^1$ & $z^2$ & $z^3$
& $z^4$ & $z^5$ & $z^6$ \\
\hline
\end{tabular}
\end{center}
}
\caption{\footnotesize 
The D3-branes solutions with waves.  
The symbols $\circ$ and $\star$ denote the directions along which 
the D3-branes and the waves extend, respectively.}
\label{D3p}
\end{table}

\subsubsection*{Simpler solutions with $k=0$}

It is worth noting a relation to other solutions by considering a special 
case of the solution with (\ref{D3p:solution-w:Eq}) and 
(\ref{D3p:solution-h:Eq}). Let us impose the conditions,   
\Eq{
\gamma_{ij}=\delta_{ij}\,,~~~u_{ab}=\delta_{ab}\,,
~~~\lap_{\Ysp}w_2\ne 0\,,~~~k=0\,.
 \label{D3p:flat2:Eq}
 }
Then we can obtain simpler solutions given by 
\cite{
Youm:1999ti, Cvetic:2000cj, Minamitsuji:2012if}
\Eqrsubl{D3p:kne0:Eq}{
h_{\rm W}(t, x, y^i, z^a)&=&a_0 + a_1\,t + a_2\,x
+\sum_{\beta=1}^{N}M_{\beta}\left(|\,\vect{y}-\vect{y}_{\beta}|^2
+L|\,\vect{z}-\vect{z}_0|^{-2}\right),
 \label{D3p:hw:Eq}\\
h(z^a)&=&\frac{L}{|\,\vect{z}-\vect{z}_0|^4}\,.
 \label{D3p:h:Eq}
}
Here $L$ is a constant parameter. 
The positions of wave and D3-branes are represented  
by $\,\vect{y}_{\beta}$ and $\vect{z}_0$\,, respectively. 
Then the distance in the two-dimensional space with $y^1$ and $y^2$ is defined as  
\Eq{
|\,\vect{y}-\,\vect{y}_\beta|
\equiv \sqrt{\left(y^1-y^1_\beta\right)^2+\left(y^2-y^2_\beta\right)^2}\,.
   \label{D3p:y:Eq}
}
Thus the solutions with (\ref{D3p:solution-w:Eq}) and (\ref{D3p:solution-h:Eq}) 
can be regarded as a generalization of the solution found in 
\cite{Youm:1999ti, Cvetic:2000cj, Minamitsuji:2012if}.

\section{Dynamical F-strings intersecting D2-branes}
\label{D2}

We shall consider a T-dual picture of the solutions presented in 
Sec.~\ref{wave}. The T-dualized solutions describe 
a brane system which consists of dynamical F-strings intersecting D2-branes. 
We also discuss collision of F-strings and the near-horizon geometries.  

\subsection{Time-dependent F-strings intersecting D2-branes}

The starting point is the D3-brane solutions with (\ref{D3p:solution-w:Eq}) 
and (\ref{D3p:solution-h:Eq})\,.  
To perform a T-duality along the $x$ direction, we have to set $a_2=0$\,. 
Hence $h_W$ does not depend on $x$\,. 

\medskip 

The T-duality relations from type IIB to type IIA are given by
\cite{Bergshoeff:1994cb, Bergshoeff:1995as, 
Breckenridge:1996tt, Costa:1996zd}
\Eqr{
&&g^{(\rm A)}_{xx}=\frac{1}{g^{(\rm B)}_{xx}}\,,~~~~
g^{(\rm A)}_{\mu\nu}=g^{(\rm B)}_{\mu\nu}
-\frac{g^{(\rm B)}_{x\mu}g^{(\rm B)}_{x\nu}
-B^{(\rm B)}_{x\mu}B^{(\rm B)}_{x\nu}}
{g^{(\rm B)}_{xx}}\,,~~~~
g^{(\rm A)}_{x\mu}=-\frac{B^{(\rm B)}_{x\mu}}{g^{(\rm B)}_{xx}}\,,\nn\\
&&\e^{2\phi_{(\rm A)}}=\frac{\e^{2\phi_{(\rm B)}}}{g^{(\rm B)}_{xx}}\,,~~~~
C_{\mu}=C_{x\mu}+\chi B_{x\mu}^{(\rm B)}\,,~~~~C_x=-\chi\,,\nn\\
&&B^{(\rm A)}_{\mu\nu}=B^{(\rm B)}_{\mu\nu}
+2\frac{B^{(\rm B)}_{x[\mu}\,g^{(\rm B)}_{\nu]x}}{g^{(\rm B)}_{xx}}\,,~~~~
B^{(\rm A)}_{x\mu}=-\frac{g^{(\rm B)}_{x\mu}}{g^{(\rm B)}_{xx}}\,,~~~~
C_{x\mu\nu}=C_{\mu\nu}
+2\frac{C_{x[\mu}\,g^{(\rm B)}_{\nu]x}}{g^{(\rm B)}_{xx}}\,,\nn\\
&&C_{\mu\nu\rho}=C_{\mu\nu\rho x}+\frac{3}{2}\left(
C_{x[\mu}\,B^{(\rm B)}_{\nu\rho]}
-B^{(\rm B)}_{x[\mu}\,C_{\nu\rho]}
-4\frac{B^{(\rm B)}_{x[\mu}\,C_{|x|\nu}g^{(\rm B)}_{\rho] x}}
{g^{(\rm B)}_{xx}}\right)\,,
   \label{t:duality:Eq}
}
where $x$ is the coordinate to which the T-duality is performed.  
The indices $\mu$\,, $\nu$ and $\rho$ denote the other coordinates.  

\medskip 

With the relations in (\ref{t:duality:Eq}), the type IIA metric is given by 
\Eqrsubl{t:solution:Eq}{
ds^2_{(\rm E)}&=& 
     h^{3/8}(z^a)h_{\rm W}^{1/4}(t, y^i, z^a)\left[
     -h^{-1}(z^a)h_{\rm W}^{-1}(t, y^i, z^a)dt^2
     +h_{\rm W}^{-1}(t, y^i, z^a)dx^2\right.\nn\\
     & &\left.+h^{-1}(z^a)\gamma_{ij}(\Ysp)dy^idy^j
     +u_{ab}(\Zsp)dz^adz^b\right],
   \label{t:metric:Eq}\\
C_{(3)}&=&h^{-1}(z^a)dt\wedge \Omega(\Ysp)\,, \quad 
B_{(2)}=\left(h_{\rm W}^{-1}(t,y^i,z^a)-1\right)dt\wedge dx\,, \nonumber \\ 
C_{(1)} &=&-\chi\,dx\,, \quad  \e^{2\phi_{(\rm A)}} = h^{1/2}(z^a)
h_{\rm W}^{-1}(t,y^i,z^a)\,. 
}
Here $ds^2_{{(\rm E)}}$ is the ten-dimensional metric with the Einstein frame. 
Then $B_{(2)}$\,, $C_{(1)}$ and $C_{(3)}$ are the 
Neveu-Schwarz-Neveu-Schwarz two-form, 
the Ramond-Ramond (RR) one-form and the RR three-form, respectively.  
The field strengths of them are defined as 
$H_{(3)} \equiv d B_{(2)}$\,, $F_{(2)} \equiv dC_{(1)}$ and $F_{(4)} 
\equiv dC_{(3)}$\,.  
The resulting metric describes a brane system composed of dynamical F-strings 
intersecting D2-branes. 
The directions along which the F-strings and the D2-branes extend are listed 
in Table \ref{D2F1}. 

\begin{table}[h]
\vspace*{0.5cm}
{\scriptsize
\begin{center}
\begin{tabular}{|c|c|c|c|c|c|c|c|c|c|c|}
\hline
Branes&0&1&2&3&4&5&6&7&8&9\\
\hline
D2 & $\circ$ && $\circ$ & $\circ$ &  &  &
&  &  &  \\
\cline{2-11}
F1 & $\circ$ & $\circ$  & & & & & & &  &  \\ 
\cline{2-11}
$x^N$ & $t$ & $x$ & $y^1$ & $y^2$ & $z^1$ & $z^2$ & $z^3$
& $z^4$ & $z^5$ & $z^6$ \\
\hline
\end{tabular}
\end{center}
}
\caption{\footnotesize 
Dynamical D2-F1 system in the metric (\ref{t:solution:Eq})\,. 
The symbol $\circ$ denotes the directions along which the world volumes 
extend. 
The brane configuration obeys the intersection rule. \label{D2F1}}
\end{table}

Let us check the equations of motion as a consistency check and consider a 
further generalization. First of all, 
the Bianchi identities for  $H_{(3)}$\,, $F_{(2)}$ and $F_{(4)}$ 
are automatically satisfied due to the T-duality relations in 
(\ref{t:duality:Eq})\,.  
Then the equation of motion for $H_{(3)}$ is decomposed into a set of 
the five equations as follows:   
\Eqrsubl{t:H3:Eq}{
&&\pd_t\pd_ih_{\rm W}=0\,,~~~~
\pd_x\pd_ih_{\rm W}=0\,,~~~~\pd_t\pd_ah_{\rm W}=0\,,
~~~~\pd_x\pd_ah_{\rm W}=0\,,
    \label{t:H3-1:Eq}\\
&&\lap_{\Ysp} h_{\rm W}+h^{-1}
\lap_{\Zsp} h_{\rm W}=0\,.
    \label{t:H3-2:Eq}
}
Note that the second and the fourth equations in (\ref{t:H3-1:Eq}) are trivially satisfied  
because there is no $x$ dependence now. 
The equation of motion for $F_{(2)}$ is reduced to 
\Eq{
k\,d\left(h\,h_{\rm W}\right)\wedge \Omega(\Ysp)\wedge \Omega(\Zsp)=0\,,
   \label{t:F2:Eq}
}
where $\Omega(\Ysp)$ and $\Omega(\Zsp)$ denote the volume forms,
defined as 
\Eqrsubl{t:volume:Eq}{
\Omega(\Ysp)&\equiv &\sqrt{\gamma}\,dy^1\wedge dy^2,
   \label{t:volume-y:Eq}\\
\Omega(\Zsp)&\equiv &\sqrt{u}\,dz^1\wedge dz^2\wedge\cdots\wedge dz^6\,. 
   \label{t:volume-z:Eq}
}

\medskip 

Now that $\pd_th_{\rm W}\ne 0$ is assumed, 
the condition (\ref{t:F2:Eq}) leads to the constraint $k=0$\,. 
This means that $F_{(2)}=0$\,. 
This is consistent with the equation of motion for $H_{(3)}$ given in 
(\ref{t:H3:Eq})\,. 
Finally, from the field equation for $F_{(4)}$\,, we obtain the following 
equation: 
\Eq{
\lap_{\Zsp} h=0\,.
   \label{t:F4:Eq}
}

\medskip

Let us suppose that $\lap_{\Ysp} h_{\rm W}=0$\,. Then, from (\ref{t:H3-2:Eq})\,, 
we obtain the relation, $\lap_{\Zsp} h_{\rm W}=0$\,. 
The solutions of $h$ and $h_{\rm W}$ are given by, respectively, 
\Eqrsubl{t:warp:Eq}{ 
\hspace{-0.8cm}h_{\rm W}(t, x, y^i, z^a)&=&a_0+a_1\,t+ a_2\,x+b_i\,y^i
+\sum_{\alpha=1}^{N}\frac{M_{\alpha}}
{2|\,\vect{z}-\,\vect{z}_{\alpha}|^4}\,,~~~~
M_{\alpha}=\frac{(2\pi l_s)^6g_s^2N_\alpha}{4\Omega_5}\,,
\label{t:hw:Eq}\\
\hspace{-0.8cm}h(z^a)&=&1+\sum_{\ell=1}^{N'}\frac{L_\ell}
{|\,\vect{z}-\,\vect{z}_\ell|^4}\,.
  \label{t:h:Eq}
  } 
Here the lengths are defined in (\ref{D3p:z:Eq})\,, and
the volume of five-sphere with the unit radius is given by
$\Omega_5$\,. The parameters $b_i$ are constant. 
Then $M_{\alpha}~ (\alpha=1\,,\cdots\,, N)$ and $L_{\ell}~ 
(\ell=1\,,\cdots\,, N')$ 
are mass parameters of F-strings and D2-branes, respectively. 
The F-strings (D2-branes) are located at $\vect{z}=\vect{z}_\alpha$ 
(~$\vect{z}_\ell$)\,. 

\medskip 

Note that we allowed the solution $h_{\rm W}$ (\ref{t:hw:Eq}) to depend on $x$ again. 
In fact, this generalization is possible. 
The second and the fourth equations in (\ref{t:H3-1:Eq}) are also satisfied. 
Thus we will discuss a T-dual picture of the D3-brane solutions with a slight generalization.

\subsection{Smearing out some of the transverse directions}

Let us next see the behavior of the solutions with (\ref{t:hw:Eq}) 
and (\ref{t:h:Eq})\,. 
First of all, the solutions are time-dependent. It is obvious from 
the expression of 
the metric. By putting $h_{\rm W}$ (\ref{t:hw:Eq}) into the metric 
(\ref{t:metric:Eq})\,, the following metric is obtained: 
\Eqr{
ds^2&=&
h^{3/8}(z^a)\left[a_0+a_1 t+a_2\,x+b_i\,y^i+w_2(z^a)\right]^{1/4}\left[
     \left\{a_0+a_1 t+a_2\,x+b_i\,y^i+w_2(z^a)\right\}^{-1}\right.\nn\\
     & &\left.\times \left\{-h^{-1}(z^a)dt^2
     +dx^2\right\}+h^{-1}(z^a)\delta_{ij}(\Ysp)dy^idy^j
     +\delta_{ab}(\Zsp)dz^adz^b\right],
   \label{t:sufrace:Eq}
}
where the function $w_2$ is defined by 
\Eq{
w_2(z^a)=\sum_{\alpha=1}^{N}\frac{M_{\alpha}}
{2|\,\vect{z}-\,\vect{z}_{\alpha}|^4}\,.
} 
The time dependence appears through the function $h_{\rm W}$\,. 
Hence the next task is to study the time evolution of the solutions carefully. 
Here we shall consider it by focusing upon collision of F-strings. 

\medskip

Here, in order to decrease the number of transverse dimensions to D2-F1 brane effectively, 
let us smear out some of the directions in the Z space.  
  
When the number of the smeared direction is given by $d_{\rm s}$\,, 
the functions $w_2$ and $h$ are rewritten as, respectively, 
\Eqrsubl{t:warp2:Eq}{
w_2(z^a)&=&\sum_{\alpha=1}^{N}\frac{M_{\alpha}}
{2|\,\vect{z}-\,\vect{z}_{\alpha}|^{4-d_{\rm s}}}\,,
\label{t:w2-2:Eq}
\\
h(z^a)&=&1+\sum_{\ell=1}^{N'}\frac{L_\ell}{|\,\vect{z}-\,\vect{z}_\ell|
^{4-d_{\rm s}}}\,.
\label{t:h-2:Eq}
}
Hereafter, we will use the smeared solutions. 

\subsection{The behavior of the solutions} 

Let us see here the asymptotic behavior of the solutions. 
The harmonic function $w_2$ dominates in the limit of 
$\,\vect{z}\rightarrow \,\vect{z}_{\alpha}$ (near a position of F-strings) 
and the time dependence can be ignored. Thus the system becomes static. 
On the other hand, in the limit of $|\,\vect{z}|\rightarrow \infty$, 
$w_2$ vanishes. Then $h_{\rm W}$ depends only on 
time $t$ in the far region from F-strings and the resulting metric is given by
\Eqr{
ds^2&=&\left(a_0+a_1 t+a_2\,x+b_i\,y^i\right)^{1/4}\left[
     \left(a_0+a_1 t+a_2\,x+b_i\,y^i\right)^{-1}\left\{-h^{-1}(z^a)dt^2
     +dx^2\right\}\right.\nn\\
     & &\left.+h^{-1}(z^a)\delta_{ij}(\Ysp)dy^idy^j
     +\delta_{ab}(\Zsp)dz^adz^b\right]\,.
   \label{t:sufrace2:Eq}
}

\medskip 

The metric becomes singular at $h_{\rm W}=0$ or $h=0$\,. Therefore 
the spacetime is regular when it is restricted inside the domain specified by the conditions,  
\Eq{
h_{\rm W}(t, x, y^i, z^a) = a_0+a_1\,t+a_2\,x+b_i\,y^i+w_2(z^a)>0\,,~~~~~
h(z^a)>0\,.
}
Here the function $w_2$ is defined in (\ref{t:w2-2:Eq}).
The spacetime cannot be extended beyond this region because
the spacetime evolves into a curvature singularity. 
Note that the regular spacetime with two F-strings  
ends up with the singular hypersurfaces. 

\medskip 

The dynamics of the spacetime depends on the signature of $a_1$\,. 
The system with $a_1>0$ has the time 
reversal one of $a_1<0$. 
In the following, we will study the case with $a_1<0$.  
Then the function $h_{\rm W}$ is positive everywhere for $t<0$ and  
the spacetime is nonsingular. 
In the limit of $t\rightarrow -\infty$, the solution is approximately 
described by 
a time-dependent uniform spacetime (apart from a position of branes, 
near which the geometry takes a cylindrical form of infinite throat).   

\medskip

Let us study the time evolution for $t>0$\,. At the initial time $t=0$\,, 
the spacetime is regular everywhere and 
has a cylindrical topology near each brane. 
As time slightly goes, a curvature singularity 
appears as $|\,\vect{z}-\,\vect{z}_{\alpha}|\rightarrow\infty$\,. 
The singular hypersurface cuts off more and more of the 
space as time goes further. 
When $t$ continues to increase, the singular hypersurface eventually splits 
and surrounds each of the F-string throats individually. 
The spatial surface is composed of two isolated throats.
For $t<0$, the time evolution of the spacetime is the time reversal of $t>0$. 

\medskip 

For any values of fixed $z^a$ in the regular domain, 
the metric (\ref{t:sufrace2:Eq}) implies that 
the overall transverse space tends to expand asymptotically like $t^{3/8}$\,. 
Thus, the solutions describe static intersecting brane systems composed of 
F-strings and D2-branes 
near the positions of the branes. On the other hand, in the far region as 
$|\,\vect{z}-\,\vect{z}_{\alpha}| \rightarrow \infty$)\,, the solutions 
approach 
FRW universes with the power law expansion $t^{3/8}$\,. The emergence of 
FRW universes is an important feature of the dynamical brane solutions.

\subsection{Collision of F-strings}

We shall argue whether two F-strings can collide or not. 
We put the two F-strings at $\vect{z}_1=(0, 0,\cdots, 0)$ 
and $\vect{z}_2=(\lambda, 0,\cdots, 0)$\,, where $\lambda$ is a constant parameter. 
In addition, we suppose that $N'$ D2-branes are sitting at a point, namely 
$z_{1}^a = \cdots = z_{N'}^a \equiv z_0^a$ 
and introduce the total mass of the $N'$ D2-branes as  
\begin{eqnarray}
L \equiv \sum_{\ell=1}^{N'}L_{\ell}\,. \label{L}
\end{eqnarray}  
It is helpful to introduce the following quantity:
\Eq{
z_{\perp}=\sqrt{\left(z^2\right)^2+\left(z^3\right)^2+\cdots 
+\left(z^{6-d_{\rm s}}\right)^2}\,.
}
Then the proper distance at $z_{\perp}=0$ between the two F-strings 
is represented by  
\Eqr{
d(t, x, y^i)&=&
\int^{\lambda}_0\! dz^1 \left(1+\frac{L}{|z^1-z^1_0|^{4-d_{\rm s}}}\right)^
{3/16}\nn\\
&&\times \left(a_0+a_1\,t+a_2\,x+b_i\,y^i+\frac{M_1}{2|z^1|^{4-d_{\rm s}}}
+\frac{M_2}{2|z^1-\lambda|^{4-d_{\rm s}}}\right)^{1/8}\,.
\label{t:distance:Eq}
}
This is a monotonically decreasing function of time. 
Here $L$ is the mass of the D2-brane. The behavior of the proper distance 
is different depending on the number of the smeared directions $d_{\rm s}$\,. 
We shall consider it for each of the values of $d_{\rm s}$ below.

\subsubsection*{The case with $d_{\rm s} \leq 4$}

\begin{figure}[tbp]
 \begin{center}
  \includegraphics[keepaspectratio, scale=.55]{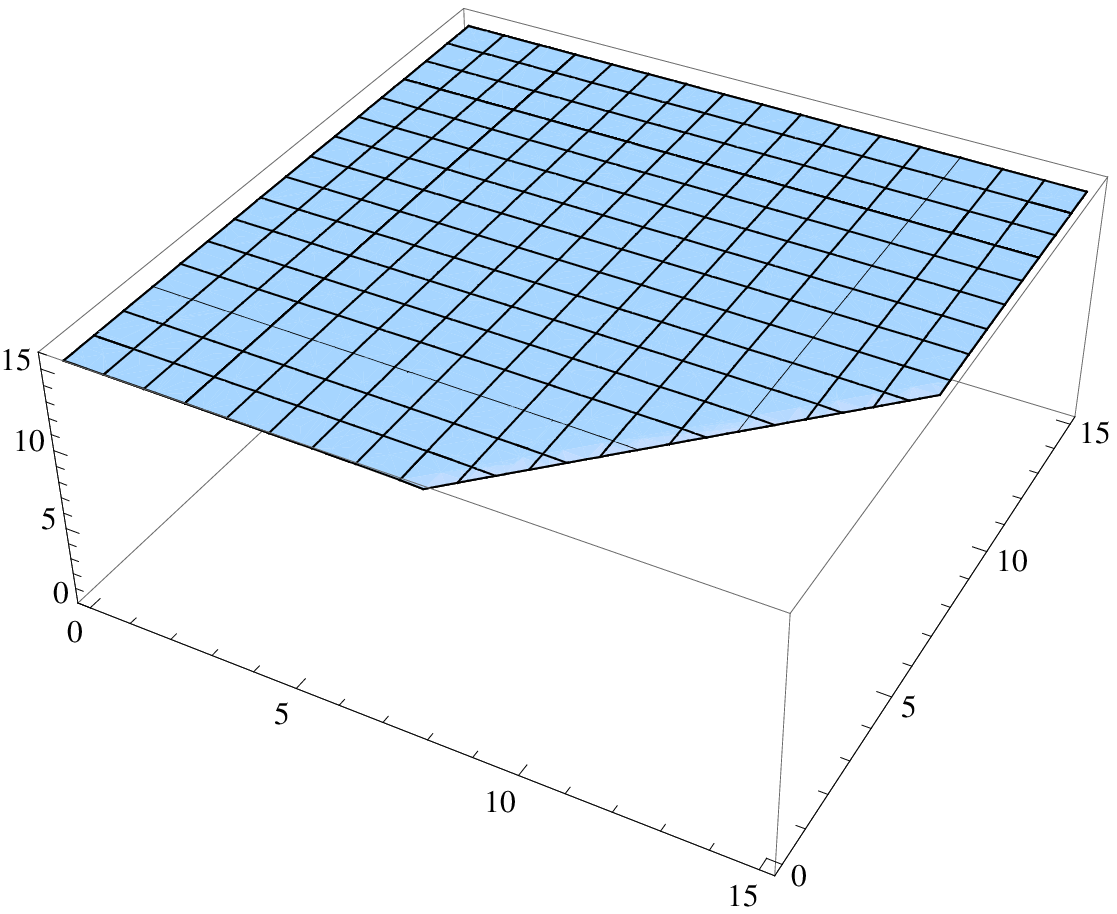}
\put(-195,100){$d(t, x)$}
\put(-100,0){$t$}
\put(5,70){$x$}
\hskip 1.5cm
\includegraphics[keepaspectratio, scale=.55]{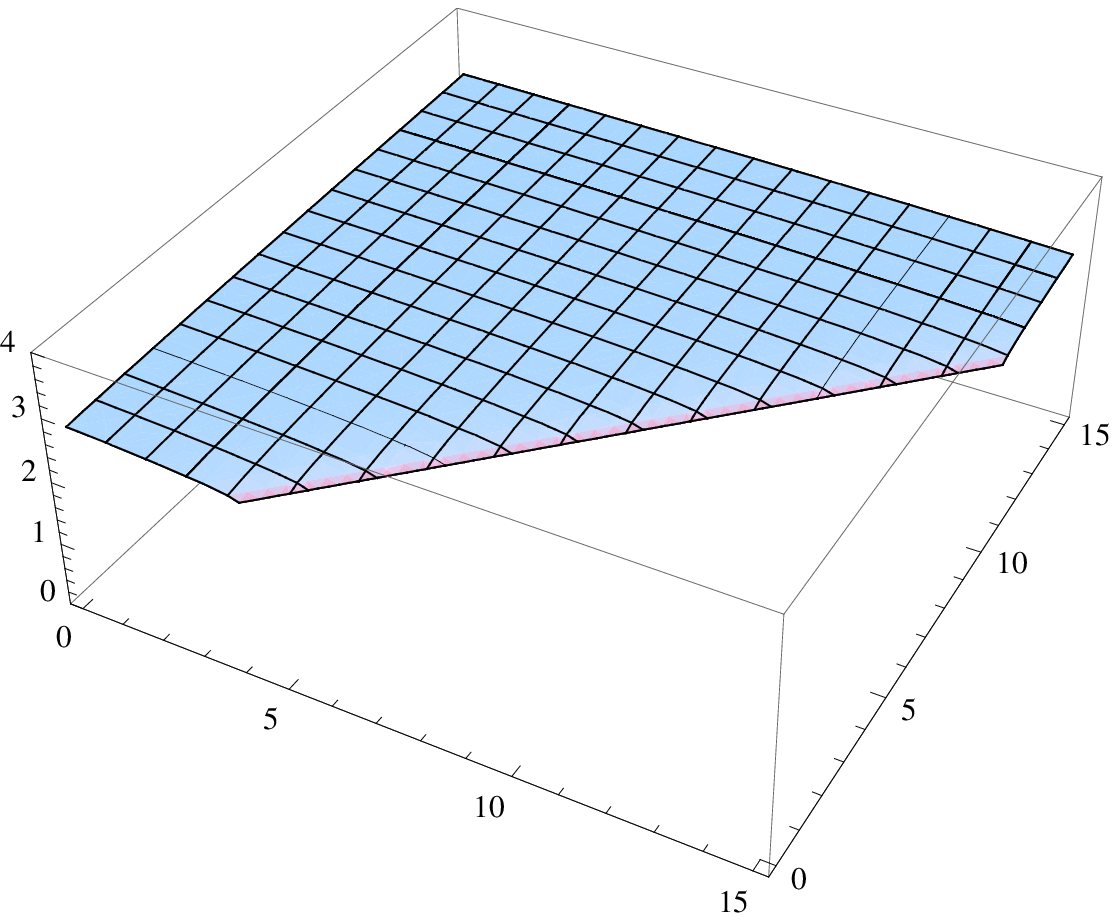}
\put(-195,100){$d(t, x)$}
\put(-100,0){$t$}
\put(5,70){$x$}
\\
 (a) \quad $d_{\rm s}=1$
\hskip 6cm 
(b) \quad $d_{\rm s}=2$~~~~~
  \caption{\footnotesize 
Two plots of the proper distance between two F-strings. 
For both cases, the two charges are identical, $M_1=M_2=L=1$ and 
the parameters are taken as $a_0=0$, $a_1=-1$, $a_2=1$, $b_i=0$, $z_0=0$, $\lambda=1$\,. 
The difference is just the value of $d_{\rm s}$\,. The result is also the same and a singularity 
develops before collision. The plots in (a) and (b) correspond to 
$d_{\rm s}=1$ and $d_{\rm s}=2$\,, 
respectively. 
  }
  \label{fig:F1}
 \end{center}
\end{figure}

Let us consider the case with $d_{\rm s} \leq 4$\,. The proper distance is plotted in Fig.~\ref{fig:F1} 
for the cases with $d_{\rm s}=1$ (left) and $d_{\rm s}=2$ (right)\,. 
Both plots mean that 
a singularity appears before the proper distance becomes zero. 
Hence the singularity between two F-strings develops before collision. 
The two F-strings approach very slowly, and then the singular hypersurface 
suddenly appears at a finite distance. After that, the spacetime splits into 
two isolated brane throats. Therefore one cannot see collision of the F-strings in these examples. 
For the other cases with $d_{\rm s}=3$ and $d_{\rm s}=4$\,, 
the result is the same.

\subsubsection*{The case with $d_{\rm s}=5$}

The next is to consider the case with $d_{\rm s}=5$\,. 
We assume that the $z^a$ directions apart from $z^1$ are smeared. 
The functions $h$ and $h_{\rm W}$ are linear in $z$\,. 
Hence the behavior of the proper distance is different from the previous case.  
The metric is given by (\ref{t:sufrace:Eq})\,. 
By choosing $z=z^1$\,, the harmonic functions $h$ and $w_2(z)$ 
are given by 
\Eqr{
w_2(z)=\frac{1}{2}\sum_{\alpha=1}^{N}M_{\alpha}
|\,z-\,z_{\alpha}|\,, \\ 
h(z)=1+\sum_{\ell=1}^{N'}L_{\ell}|\,z-\,z_{\ell}|\,.
    \label{t:w2h:Eq}
}

\medskip 

Let us consider the solutions in the case that one F-string charge $M_1$ is located 
at $z=0$ and the other $M_2$ at $z=\lambda$\,. 
The proper distance between the two F-strings is represented by
\Eqr{
d(t, x, y^i)&=&\int^{\lambda}_0 dz \left[a_0+a_1t+a_2\,x+b_i\,y^i
  +\frac{1}{2}\left(M_1|z|+M_2|z-\lambda|\right)\right]^{1/8}\nn\\
 &&\times\left(1+ L|z-z_0|\right)^{3/16}\,.
  \label{t:length:Eq}
}
For $a_1<0$, the proper length decreases with time. 
In the case that $M_1\ne M_2$\,, a singularity appears again at a certain finite time $t=t_{\rm c}$\,, 
while the distance is still finite. Here $t_{\rm c}$ is defined as 
\[
t_{\rm c}\equiv-[2a_0+2a_2\,x+2b_i\,y^i+M_1|z|+M_2|z-\lambda|]/2a_1>0\,. 
\] 
This is the same result as the case with $d_{\rm s}\leq 4$\,. 

\medskip 

However, in the same charge case that $M_1=M_2=M$\,, 
the distance indeed vanishes at a certain finite time $t=t_{\rm c}$\,, 
where $t_{\rm c}$ is defined as   
\[
t_{\rm c} \equiv -(2a_0+2a_2\,x+2b_i\,y^i+M\lambda)/2a_1\,.
\] 
Hence two F-strings can collide completely. 

\medskip 

\begin{figure}[tbp]
 \begin{center}
  \includegraphics[keepaspectratio, scale=.55]{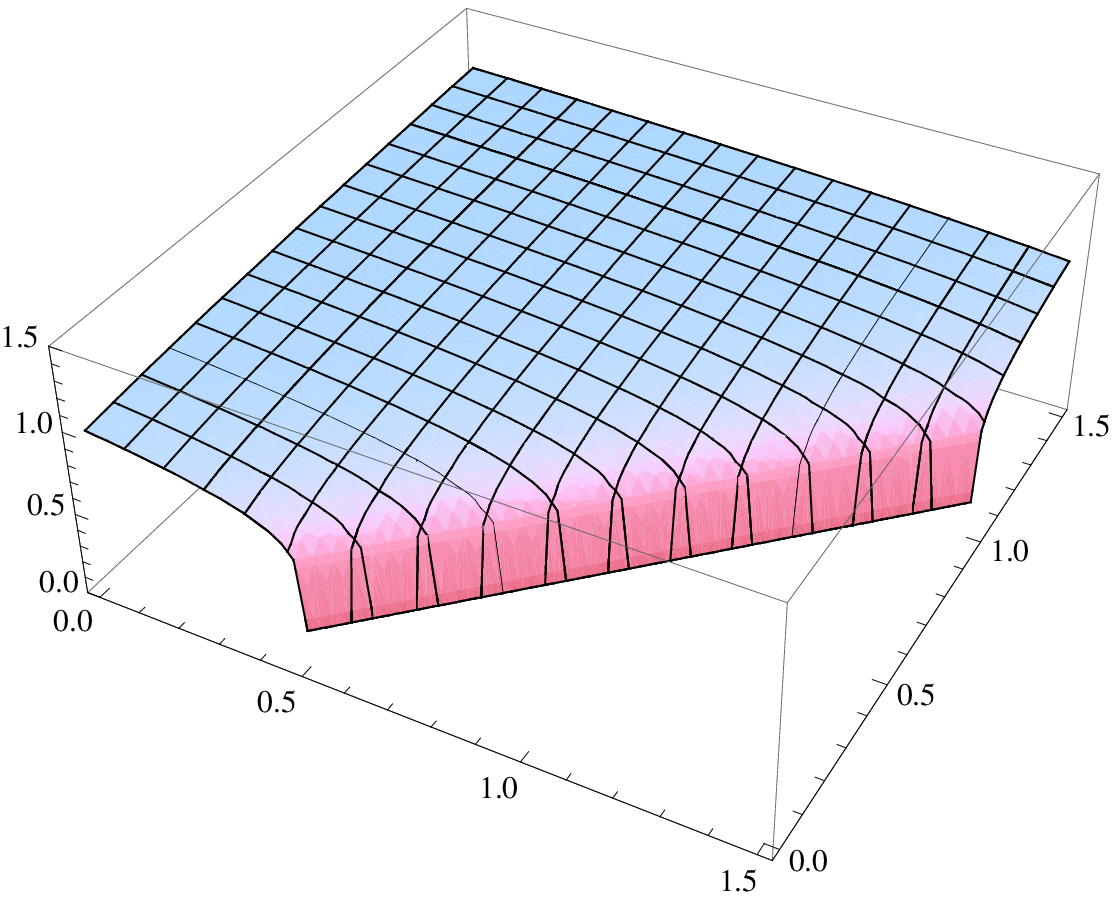}
\put(-190,100){$d(t, x)$}
\put(-100,0){$t$}
\put(5,70){$x$}
\hskip 1.5cm
\includegraphics[keepaspectratio, scale=.55]{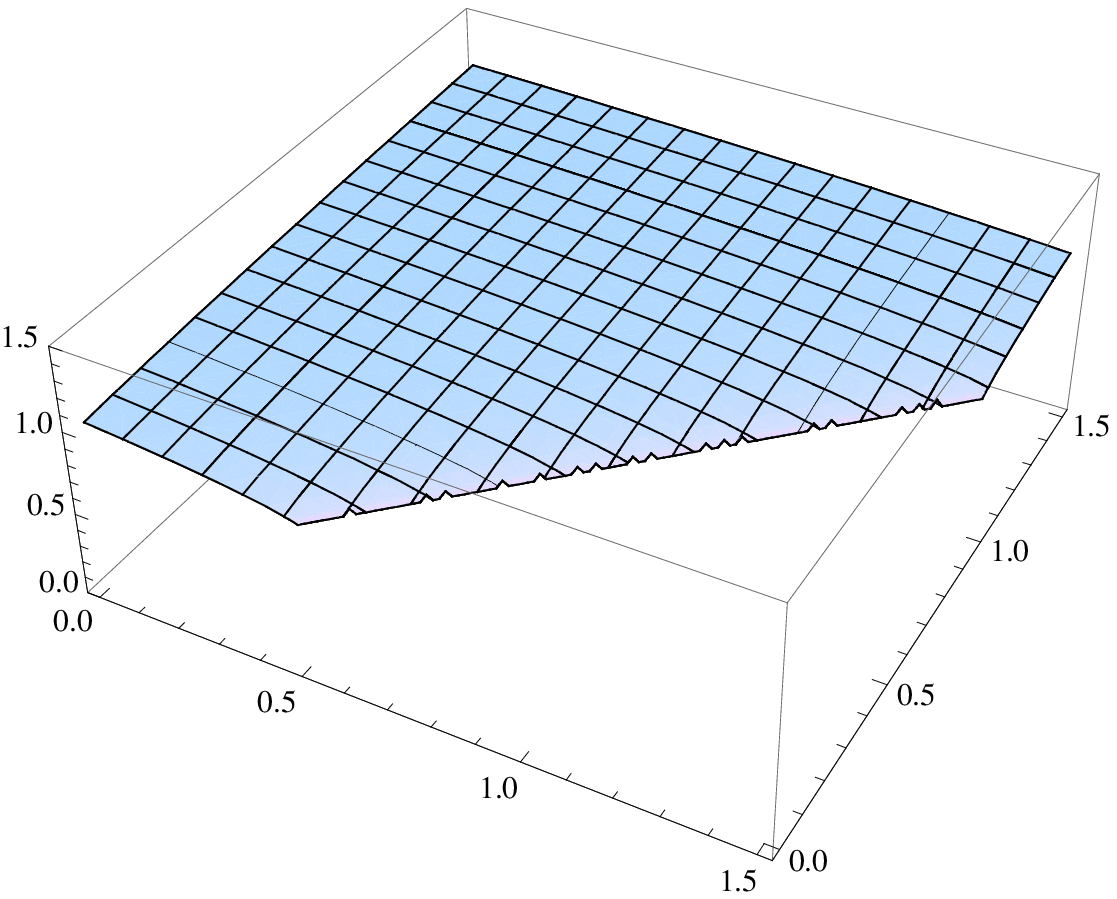}
\put(-190,100){$d(t, x)$}
\put(-100,0){$t$}
\put(5,70){$x$}
\\
 (a) \quad $M_1=M_2$
\hskip 5cm (b) \quad $M_1\neq M_2$
  \caption{\footnotesize 
Two plots of the proper distance between the F-strings. For both cases, 
the parameters are set as $a_0=0$, $a_1=-1$, $a_2=1$, $b_i=0$, $z_0=0$, $\lambda=1$ 
and $L=1$\,. 
The difference is the values of the charges.  
For the plot in (a)\,,  
$M_1=M_2=1$\,. Then the complete collision occurs at finite time $t=t_{\rm c}$\,. 
For the plot in (b)\,, $M_1=2$ and $M_2=1$\,. 
Then a singularity suddenly appears before collision again. 
  }
  \label{fig:F1-D2}
 \end{center}
\end{figure}

By using $t_{\rm c}$\,, the proper distance is expressed as 
\Eqr{
d(t, x, y^i)=\frac{16}{19L}\left[-1+\left(1+\lambda L\right)^{19/16}\right]
\left[a_1(t-t_{\rm c})\right]^{1/8}\,.
} 
For the values as 
$a_0=0$\,, $a_1=-1$\,, $a_2=1$\,, $b_i=0$\,, $z_0=0$\,, $\lambda=1$ and 
$L=1$\,, the proper distance $d(t, x)$ is plotted in Fig.~\ref{fig:F1-D2} 
for the two cases (a) the same charges $M_1=M_2=1$ and (b) different 
charges $M_1=2$, $M_2=1$\,. 
In the case (a) the two F-strings can collide completely, but in the case 
(b) a singularity appears before collision, as we have already seen 
analytically.

\subsection{Near-horizon geometry}

Finally we shall see the near-horizon geometry of the solutions. 
When all of the F-strings and D2-branes are located at the origin of the Z
spaces, the solutions are rewritten as  
\Eqr{
h_{\rm W}(t, x, y^i, r)&=&a_0+a_1\,t +a_2\,x+b_i\,y^i+\frac{M_{\rm W}}{2r^4}\,,
\label{nh:hw:Eq}\\
h(r)&=&1+\frac{L}{r^4}\,, \qquad r^2 \equiv \delta_{ab}\, z^a z^b\,. 
  \label{nh:h:Eq}
}
Here $L$ is introduced in (\ref{L}) and $M_{\rm W}$ is the total mass of F-strings  
\[
M_{\rm W} \equiv \sum_{\alpha=1}^N M_{\alpha}\,. 
\]
In the near-horizon region $r\rightarrow 0$\,, 
the dependence on $t$ and $y^i$ in (\ref{nh:hw:Eq}) is negligible 
and hence the metric is reduced to the following form:  
\Eq{
ds^2=L^{-\frac{5}{8}}\,\left(\frac{M_{\rm W}}{2}\right)
^{-\frac{3}{4}}\,\tilde{r}^{-\frac{1}{6}}\left[ds^2_{\rm AdS_2}
+\left(L\,dx^2+\frac{M_{\rm W}}{2}\,
(dy^i)^2\right)\tilde{r}^{\frac{2}{3}}
+\frac{M_{\rm W}L}{2}d\Omega^2_{(5)}\right]\,, \label{resulting}
}
where $r^3=\tilde{r}$ has been performed. 
The line elements of a two-dimensional AdS space (AdS$_2$) 
and a five-sphere with the unit radius ($S^5$) 
are given by $ds^2_{\rm AdS_2}$ and $d\Omega^2_{(5)}$\,, respectively. 
Thus the metric (\ref{resulting}) describes a warped product of AdS${}_2$ 
and $S^5$\,.

\subsubsection*{A simpler case}

Let us consider a simpler case that the spacetime metric and 
$h_{\rm W}$ satisfy (\ref{D3p:flat2:Eq})\,.  
It is convenient to perform a coordinate transformation for $y^i$ and $z^a$ 
as follows \cite{Cvetic:2000cj}:  
\Eq{
y^1=\frac{1}{r}\cos\theta\cos\alpha\,, \quad 
y^2=\frac{1}{r}\sin\theta\cos\alpha\,, \quad 
z^a=\frac{rL^{1/2}}{\sin\alpha}\mu^a\,.
   \label{nh:trans:Eq}
}
Here the unit vector $\mu^a$ parametrizes $S^5$ and satisfies the 
following conditions: 
\Eq{
\mu_a\mu^a=1\,,~~~~~~d\Omega^2_{(5)}=d\mu_ad\mu^a\,.
}
In terms of the new coordinates (\ref{nh:trans:Eq})\,, 
$h_{\rm W}$ and $h$ are rewritten as 
\Eqrsubl{nh:solution:Eq}{
h_{\rm W}(t, x, r)&=&a_0+a_1\,t +a_2\,x+\frac{M_{\rm W}}{r^2},
 \label{nh:hw2:Eq}\\
h(r)&=&\frac{\sin^4\alpha}{L\,r^4}\,,
 \label{nh:h2:Eq}
}
where $M_{\rm W}$ is a constant parameter and 
the F-strings are assumed to be at the origin of the Y space and the Z space.

\medskip 

In the near-horizon region $r\rightarrow 0$, the metric is described by 
\cite{Minamitsuji:2012if, Cvetic:2000cj}
\Eqr{
ds^2&=&M_{\rm W}^{1/4}\,L^{5/8}\,(\sin\alpha)^{-5/2}
\left[-\frac{r^4}{M_{\rm W}}dt^2+\frac{dr^2}{r^2}+d\alpha^2
+\cos^2\alpha\,d\theta^2\right.\nn\\
&&\left.+\sin^2\alpha\,d\Omega_{(5)}^2
+\left(M_{\rm W}\,L\right)^{-1}\sin^4\alpha\left(dx\right)^2\right].
    \label{nh:metric2:Eq}
}
This is a warped product of a two-dimensional Lifshitz spacetime 
with an eight-dimensional internal space. Note that the Lifshitz part can be rewritten as 
an AdS$_2$ with a coordinate transformation $r^2 =u$\,.

\section{Cosmological aspects of the solutions} 
\label{cos}

Let us consider cosmological aspects of the solution describing dynamical F-strings intersecting D2-branes. 
We first study the time dependence of the scale factors in the solutions. Then, by compactifying the extra directions, 
our dynamical Universe is discussed. 

\subsection{The scale factors} 

First of all, let us see the scale factors in the solutions. 
It is helpful to introduce a new time coordinate $\tau$ as 
\Eq{
\left(\frac{\tau}{\tau_0}\right) \equiv \left(a_1t\right)^{5/8}\,,~~~~
\tau_0 \equiv \frac{8}{5a_1}\,,
}
where $a_1>0$ is assumed for simplicity. 
The ten-dimensional metric is thus given by
\Eqr{
&&\hspace{-0.2cm}ds^2_{10}=h^{-\frac{5}{8}}
\left[1+\left(\frac{\tau}{\tau_0}\right)^{-\frac{8}{5}}W(x, y^i, z^a)
\right]^{-\frac{3}{4}}
\left[-d\tau^2
+\left\{1+\left(\frac{\tau}{\tau_0}\right)^{-\frac{8}{5}}
W(x, y^i, z^a)\right\}
\left(\frac{\tau}{\tau_0}\right)^{\frac{2}{5}}
\right.\nn\\
&&\left.\hspace{1cm}\times
\left\{\delta_{ij}(\tilde{\Ysp})dy^idy^j
+h\left(\delta_{\mu\nu}dz^{\mu}dz^{\nu}
+\delta_{mn}(\tilde{\Zsp})dz^mdz^n\right)\right\}
+\left(\frac{\tau}{\tau_0}\right)^{-\frac{6}{5}}hdx^2\right],
   \label{cos:metric3:Eq}
}
where $W(x, y^i, z^a)$ is defined by 
\Eq{
W(x, y^i, z^a) \equiv a_0+a_2\,x+b_i\,y^i
+\sum_{\alpha=1}^{N}\frac{M_{\alpha}}
{2|\,\vect{z}-\,\vect{z}_{\alpha}|^{4-d_{\rm s}}}\,.
}
Recall that $d_{\rm s}$ is the number of smeared dimensions and 
should satisfy $0\le d_{\rm s}\le 5$\,. Here we assume that 
at least one direction of $z^a~(a=1,\ldots,6)$ is not smeared 
in order to specify the location of our Universe in the transverse space.

\medskip 

Let us consider a simplified case that $(\tau/\tau_0)^{-8/5}W=0$\,. 
This geometry is realized 
in the limit $\tau\rightarrow\infty$\,. 
Then the scale factor of the four-dimensional space is 
proportional to $\tau^{1/5}$\,. For the remaining spaces $\tilde{\Xsp}$ and 
$\tilde{\Ysp}$\,, 
it is proportional to $\tau^{-3/5}$ and $\tau^{1/5}$, respectively. 
Thus, unfortunately, the solutions do not to give a realistic universe such 
as an accelerating expansion, a matter dominated era or a radiation dominated 
era \cite{Burgess:2003mk, Tolley:2006ht, Anchordoqui:2007sb, Tolley:2007et}. 
However, there is a possibility that appropriate compactification and smearing 
of the extra directions 
may lead to a realistic expansion. In the next subsection we will discuss 
this possibility.

\subsection{Classifying dynamical universes}

The next is to consider some compactification and smearing of the extra 
directions of the solutions. 
First of all, our Universe has to be specified in the solutions with ten 
directions $t,~x,~y^i,~z^a~(i=1,2,~a=1,\ldots,6)$\,. The time direction 
is identified with $t$\,. 
The remaining task is to identify the three spatial directions from the 
other coordinates. 

\medskip 

In the usual approach such as the brane-world scenario, 
three spatial directions are supposed to be on the branes.  
In the present case, however, it does not work because the spatial directions are specified with $x$, $y^1$ and $y^2$ 
and then the space is not isotropic from the expression of the metric.  
Thus we have to look for another way to realize an isotropic 
and homogeneous three-dimensional space in the present solutions.  

\medskip 

The remaining choice is to take the three-dimensional 
from the overall transverse space with $z^a$. 
That is, our Universe is spanned by $t$, $z^2$, $z^3$ and $z^4$\,, 
for example. The $z^1$ direction is preserved to measure the position 
of our Universe in the overall transverse space. 
Note that the metric depends on $z^a$ explicitly. Hence we have to smear 
out $z^2$\,, $z^3$ and $z^4$ so as to define our Universe. 
Thus the number of the smeared directions 
$d_{\rm s}$ should satisfy the bound $3 \leq d_{\rm s} \leq 5$\,. 
The remaining degrees of freedom to be smeared out 
are $z^5$ and $z^6$\,. 

\medskip 

Now our Universe is described by the coordinates $(t, z^2, z^3, z^4)$\,.  
Hence it is convenient to decompose the ten-dimensional metric of 
the solutions into the following form: 
\Eq{
ds^2_{10}= ds_4^2(\tilde{\Msp})
+ds^2(\tilde{\Xsp})+ds^2(\tilde{\Ysp})+ds^2(\tilde{\Zsp})\,,
    \label{cos:metric:Eq}
}
where each part of the metric is given by 
\Eqrsubl{cos:metric2:Eq}{
ds^2_4(\tilde{\Msp})&=&h^{3/8}(z^a)\,h_{\rm W}^{1/4}(t, x, y^i, z^a)\,\left[
     -h^{-1}(z^a)\,h_{\rm W}^{-1}(t, x, y^i, z^a)\,dt^2+\delta_{\mu\nu}
     dz^{\mu}dz^{\nu}\right],\\
ds^2(\tilde{\Xsp})&=&h^{3/8}(z^a)\,h_{\rm W}^{-3/4}(t, x, y^i, z^a)\,dx^2\,,\\
ds^2(\tilde{\Ysp})&=&h^{-5/8}(z^a)\,h_{\rm W}^{1/4}(t, x, y^i, z^a)\,
                     \delta_{ij}(\tilde{\Ysp})\,dy^idy^j,\\
ds^2(\tilde{\Zsp})&=&h^{3/8}(z^a)\,h_{\rm W}^{1/4}(t, x, y^i, z^a)\,\delta_{mn}
(\tilde{\Zsp})\,dz^mdz^n\,.
}
Here $ds^2_4(\tilde{\Msp})$ is the metric of the four-dimensional spacetime 
with $t,~z^{\mu}$~$(\mu=2, 3, 4)$\,. 
The space $\tilde{\rm X}$ is one-dimensional with $x$\,.  
The space $\tilde{\Ysp}$ is described by $y^i~(i=1,2)$ and  
and the space $\tilde{\Zsp}$ is three-dimensional with $z^m~(m=1, 5 ,6)$\,.  
Then $\delta_{\mu\nu}$, $\delta_{ij}$, $\delta_{mn}$ are 
the three-, two- and three-dimensional Euclidean metrics, respectively. 

\medskip 

The next task is to derive the lower-dimensional effective theory by 
compactifying the extra directions.  
It is necessary to take that $a_0=a_2=b_i=0$ in (\ref{cos:metric3:Eq}) 
to compactify the $x$ and $y^i$ directions.  
To find a realistic universe, let us compactify 
$d=\left(d_{\tilde{\Xsp}}+d_{\tilde{\Ysp}}
+d_{\tilde{\Zsp}}\right)$-dimensional space to be a $d$-dimensional torus, 
where $d_{\tilde{\Xsp}}$\,, $d_{\tilde{\Ysp}}$ and $d_{\tilde{\Zsp}}$ 
are the compactified dimensions for the direction of $x$, $\tilde{\Ysp}$ 
and $\tilde{\Zsp}$ spaces, respectively. 
The range of $d_{\tilde{\Zsp}}$ is given by $0\le d_{\tilde{\Zsp}}\le 2$ 
because the $z^1$ direction is preserved to measure the position of the 
Universe in the overall transverse space. 
That is, the $z^5$ and $z^6$ directions may be compactified, where the direction to be compactified has to be smeared out 
before the compactification. On the other hand, for the compactification of  the $x$ and $y^i$ directions, 
the smearing is not needed because we have set that $a_0=a_2=b_i=0$\,.

\medskip

We shall classify the compactifications of the solutions depending 
on the variety of the internal space. 
The remaining noncompact space is referred to as the external space. 
Then the metric (\ref{cos:metric:Eq}) is recast into the following form: 
\Eq{
ds^2_{10}=ds^2_{\rm e}+ds^2_{\rm i}\,,
   \label{cos:metric4:Eq}
}
where $ds^2_{\rm e}$ is the metric of $(10-d)$-dimensional external spacetime 
and $ds^2_{\rm i}$ is the metric of compactified dimensions. 

\medskip 

Recall here that our four-dimensional spacetime should be described in the Einstein frame. 
The Einstein frame metric is obtained by multiplying the extra factor and 
the resulting metric, which leads to new exponents, is given by 
\Eqr{
\hspace{-0.4cm}d\bar{s}^2_{\rm e}&=&h^{\xi}
\left[1+\left(\frac{\tau}{\tau_0}\right)^{-\frac{2}{\beta+2}}W(z^a)
\right]^{\beta}
\left[-d\tau^2
+\left\{1+\left(\frac{\tau}{\tau_0}\right)^{-\frac{2}{\beta+2}}W(z^a)\right\}
\left(\frac{\tau}{\tau_0}\right)^{\frac{2\left(\beta+1\right)}{\beta+2}}
\right.\nn\\
&&\left.\hspace{-0.2cm}\times
\left\{\delta_{ij}(\tilde{\Ysp'})dy^idy^j
+h\left(\delta_{\mu\nu}dz^{\mu}dz^{\nu}
+\delta_{mn}(\tilde{\Zsp'})dz^mdz^n\right)\right\}
+\left(\frac{\tau}{\tau_0}\right)^
{\frac{2\beta}{\beta+2}}hd{x'}^2\right],
   \label{cos:Emetric:Eq}
}
where $d\bar{s}^2_{\rm e}$ is the $(10-d)$-dimensional metric in 
the Einstein frame. Here $x'$ and ${\Ysp}'$ are the coordinates of 
$(1-d_{\tilde{\Xsp}})$- and $(2-d_{\tilde{\Ysp}})$-dimensional relative transverse spaces, respectively.  
Finally ${\Zsp}'$ denotes $(3-d_{\tilde{\Zsp}})$-dimensional spaces. 

\medskip 

The constant parameters $\xi$, $\beta$, $\tau_0$ and the cosmic time $\tau$ are defined as 
\Eqrsubl{cos:pa:Eq}{
&&\xi \equiv \frac{1}{8-d}\left(d-d_{\tilde{\Ysp}}-5\right),~~~~
\beta \equiv \frac{1}{8-d}\left(d-d_{\tilde{\Xsp}}-6\right),\\
&&\frac{\tau}{\tau_0} \equiv \left(a_1t\right)^{(\beta+2)/2},~~~~
\tau_0 \equiv \frac{2}{\left(\beta+2\right)a_1}\,.
}
The power of  $\tilde{M}$ of the solutions is given by 
\Eq{
\lambda_{\rm E} \equiv \frac{\beta+1}{\beta+2}<1 \qquad {\rm for}~~8>d ~~\mbox{and}~~ d>0\,.
   \label{cos:power:Eq}
} 
In Table \ref{ta:power}, we list the power exponent of the fastest 
expansion of our $(10-d)$-dimensional Universe in the Einstein frame. 
Unfortunately, every exponent in Table \ref{ta:power} is so small that 
the solutions do not give rise to our expanding Universe. 
To realize a realistic cosmological model such as in the inflationary 
scenario, it would be necessary to add some new ingredients.

\begin{table}[tbp]
\begin{center}
{
\begin{tabular}{|c||c|c||c|}
\hline
$10-d$  & External spacetime & Internal space &
$\lambda_{\rm E}$
\\
\hline
\hline
 9& $\tilde{\Msp}$ \& $\tilde{\Xsp}'$ \& $\tilde{\Ysp}'$ \& 
 $\tilde{\Zsp}'$ & 
$(d_{\tilde{\Xsp}}, d_{\tilde{\Ysp}}, d_{\tilde{\Zsp}})=(0, 0, 1)$ &2/9
\\
\hline
9& $\tilde{\Msp}$ \& $\tilde{\Xsp}'$ \& $\tilde{\Ysp}'$ \& 
 $\tilde{\Zsp}'$ & 
$(d_{\tilde{\Xsp}}, d_{\tilde{\Ysp}}, d_{\tilde{\Zsp}})=(0, 1, 0)$ &2/9
\\
\hline
9& $\tilde{\Msp}$  \& $\tilde{\Ysp}'$ \& 
 $\tilde{\Zsp}'$ & 
$(d_{\tilde{\Xsp}}, d_{\tilde{\Ysp}}, d_{\tilde{\Zsp}})=(1, 0, 0)$ &1/8
\\
\hline
8& $\tilde{\Msp}$ \& $\tilde{\Xsp}'$ \& $\tilde{\Ysp}'$ \& $\tilde{\Zsp}'$ & 
$(d_{\tilde{\Xsp}}, d_{\tilde{\Ysp}}, d_{\tilde{\Zsp}})=(0, 0, 2)$ &1/4
\\
\hline
8& $\tilde{\Msp}$ \& $\tilde{\Xsp}'$ \& $\tilde{\Ysp}'$ \& 
 $\tilde{\Zsp}'$ & 
$(d_{\tilde{\Xsp}}, d_{\tilde{\Ysp}}, d_{\tilde{\Zsp}})=(0, 1, 1)$ &1/4
\\
\hline
8& $\tilde{\Msp}$ \& $\tilde{\Ysp}'$ \&  $\tilde{\Zsp}'$ & 
$(d_{\tilde{\Xsp}}, d_{\tilde{\Ysp}}, d_{\tilde{\Zsp}})=(1, 0, 1)$ &1/7
\\
\hline
8& $\tilde{\Msp}$ \& $\tilde{\Xsp}'$ \& $\tilde{\Zsp}'$ & 
$(d_{\tilde{\Xsp}}, d_{\tilde{\Ysp}}, d_{\tilde{\Zsp}})=(0, 2, 0)$ &1/4
\\
\hline
8& $\tilde{\Msp}$ \& $\tilde{\Ysp}'$ \& $\tilde{\Zsp}'$ & 
$(d_{\tilde{\Xsp}}, d_{\tilde{\Ysp}}, d_{\tilde{\Zsp}})=(1, 1, 0)$ &1/7
\\
\hline
7& $\tilde{\Msp}$ \& $\tilde{\Xsp}'$ \& $\tilde{\Ysp}'$ \& $\tilde{\Zsp}'$ & 
$(d_{\tilde{\Xsp}}, d_{\tilde{\Ysp}}, d_{\tilde{\Zsp}})=(0, 1, 2)$ &2/7
\\
\hline
7& $\tilde{\Msp}$ \& $\tilde{\Ysp}'$ \& $\tilde{\Zsp}'$ & 
$(d_{\tilde{\Xsp}}, d_{\tilde{\Ysp}}, d_{\tilde{\Zsp}})=(1, 0, 2)$ &1/6
\\
\hline
7& $\tilde{\Msp}$ \& $\tilde{\Xsp}'$ \& $\tilde{\Zsp}'$ & 
$(d_{\tilde{\Xsp}}, d_{\tilde{\Ysp}}, d_{\tilde{\Zsp}})=(0, 2, 1)$ &2/7
\\
\hline
7& $\tilde{\Msp}$ \& $\tilde{\Ysp}'$ \& $\tilde{\Zsp}'$ & 
$(d_{\tilde{\Xsp}}, d_{\tilde{\Ysp}}, d_{\tilde{\Zsp}})=(1, 1, 1)$ &1/6
\\
\hline
7& $\tilde{\Msp}$ \& $\tilde{\Zsp}'$ & 
$(d_{\tilde{\Xsp}}, d_{\tilde{\Ysp}}, d_{\tilde{\Zsp}})=(1, 2, 0)$ &1/6
\\
\hline
6& $\tilde{\Msp}$ \& $\tilde{\Xsp}'$ \& $\tilde{\Zsp}'$ & 
$(d_{\tilde{\Xsp}}, d_{\tilde{\Ysp}}, d_{\tilde{\Zsp}})=(0, 2, 2)$ &1/3
\\
\hline
6& $\tilde{\Msp}$ \& $\tilde{\Ysp}'$ \& $\tilde{\Zsp}'$ & 
$(d_{\tilde{\Xsp}}, d_{\tilde{\Ysp}}, d_{\tilde{\Zsp}})=(1, 1, 2)$ &1/5
\\
\hline
5& $\tilde{\Msp}$ \& $\tilde{\Zsp}'$ & 
$(d_{\tilde{\Xsp}}, d_{\tilde{\Ysp}}, d_{\tilde{\Zsp}})=(1, 2, 2)$ &1/4
\\
\hline
\end{tabular}
}
\caption{\footnotesize 
The classification of compactifications of the transverse directions and the maximum 
value of the power exponent $\lambda_{\rm E}$ for $\tilde{M}$\,. The exponent is measured 
in the Einstein frame. 
}
\label{ta:power}
\end{center}
\end{table}

\section{Conclusion and discussion}
\label{Conclusion}

We have constructed gravitational solutions composed of dynamical 
F-strings intersecting D2-branes in type IIA supergravity and studied the 
time evolution focusing upon 
the collision of F-strings. We also have discussed some applications to 
cosmology.

\medskip 

First, we have obtained time-dependent multicenter D3-brane solutions 
with multiple waves in type IIB supergravity by generalizing 
the solutions found in \cite{Uzawa:2013koa}. 
Then we have obtained the solutions composed of dynamical F-strings 
intersecting D2 branes by using a T-duality. 
The solutions can also be obtained from supersymmetric static solutions  
by replacing a constant in the harmonic function with a linear function 
of time. 
This procedure is applicable universally to construct a class of 
time-dependent brane solutions. 

\medskip 

It is worth commenting on the time dependence of the solutions. 
The solutions admit F-strings to be dynamical, but D2-branes have 
to be static. The origin of this restriction is 
that the ansatz we imposed is too restrictive. In fact, in recent 
works on similar systems, both harmonic functions may depend on time 
\cite{Minamitsuji:2011jt, Maeda:2012xb}. 
We have also discussed collision of F-strings. Two F-strings approach each other with time. 
In general, a singularity appears before collision. An exceptional case is 
that five directions in the overall transverse space 
are smeared out. 

\medskip

The dynamical solutions give rise to a singularity when 
$h_{\rm W}(t, x, y^i, z^a)=0$\,.
The appearance of the singularity indicates the two possibilities:  
(1) the ansatz is simply wrong or (2) just the supergravity approximation is 
not valid and it is necessary to include some stringy corrections from
 $\alpha'$ or $g_s$\,. 
It is difficult to decide which of them applies to the present case 
only from the supergravity perspective. It would be interesting to consider 
some generalizations to include stringy corrections.

\medskip 

We have also argued cosmological models obtained from the solutions 
by compactifying the internal space. 
The resulting FRW universes exhibit a power-law expansion. 
However, the power of the scale factor is less than $1/2$ and hence 
it is too small to realize a realistic expansion. 
To obtain a realistic expanding universe such as an 
inflationary scenario or matter or a radiation dominated era, 
we have to include additional matter fields. Thus it is interesting 
to try to construct a cosmological model 
by studying a more complicated setup. 

\medskip 

After all, the solutions constructed here do not give a realistic cosmological model. 
However, we believe that more elaborated construction along this line 
would open a new window to realize 
realistic cosmological expansions in the context of string theory.

\subsection*{Acknowledgments}

This work was also supported in part by the Grant-in-Aid 
for the Global COE Program ``The Next Generation of Physics, Spun 
from Universality and Emergence'' from MEXT, Japan.


\begin{thebibliography}{99}

\bibitem{GLP}
  G.~W.~Gibbons, H.~Lu and C.~N.~Pope,
  ``Brane worlds in collision,''
  Phys.\ Rev.\ Lett.\  {\bf 94} (2005) 131602
  [arXiv:hep-th/0501117].
  
\bibitem{CCGLP}
  W.~Chen, Z.~W.~Chong, G.~W.~Gibbons, H.~Lu and C.~N.~Pope,
  ``Horava-Witten stability: Eppur si muove,''
  Nucl.\ Phys.\  B {\bf 732} (2006) 118
  [arXiv:hep-th/0502077].

\bibitem{KU1}
  H.~Kodama and K.~Uzawa,
  ``Moduli instability in warped compactifications of the type IIB
  supergravity,''
  JHEP {\bf 0507} (2005) 061
  [arXiv:hep-th/0504193].

\bibitem{KU2}
  H.~Kodama and K.~Uzawa,
  ``Comments on the four-dimensional effective theory for warped 
  compactification,''
  JHEP {\bf 0603} (2006) 053
  [hep-th/0512104].
  
\bibitem{BSU}
  P.~Binetruy, M.~Sasaki and K.~Uzawa,
  ``Dynamical D4-D8 and D3-D7 branes in supergravity,''
  Phys.\ Rev.\  D {\bf 80} (2009) 026001
  [arXiv:0712.3615 [hep-th]].

%\cite{Maeda:2009tq}
\bibitem{Maeda:2009tq}
  K.~i.~Maeda, N.~Ohta, M.~Tanabe and R.~Wakebe,
  ``Supersymmetric Intersecting Branes in Time-dependent Backgrounds,''
  JHEP {\bf 0906} (2009) 036
  [arXiv:0903.3298 [hep-th]].
  
\bibitem{MOU}
  K.~i.~Maeda, N.~Ohta and K.~Uzawa,
  ``Dynamics of intersecting brane systems -- Classification and their
  applications --,''
  JHEP {\bf 0906} (2009) 051
  [arXiv:0903.5483 [hep-th]].

%\cite{Gibbons:2009dr}
\bibitem{Gibbons:2009dr}
  G.~W.~Gibbons and K.~i.~Maeda,
  ``Black Holes in an Expanding Universe,''
  Phys.\ Rev.\ Lett.\  {\bf 104} (2010) 131101
  [arXiv:0912.2809 [gr-qc]].

%\cite{Maeda:2009ds}
\bibitem{Maeda:2009ds}
  K.~i.~Maeda and M.~Nozawa,
  ``Black Hole in the Expanding Universe from Intersecting Branes,''
  Phys.\ Rev.\  D {\bf 81} (2010) 044017
  [arXiv:0912.2811 [hep-th]].

%\cite{Maeda:2010ja}
\bibitem{Maeda:2010ja}
  K.~i.~Maeda and M.~Nozawa,
  ``Black Hole in the Expanding Universe with Arbitrary Power-Law Expansion,''
  Phys.\ Rev.\  D {\bf 81} (2010) 124038
  [arXiv:1003.2849 [gr-qc]].

%\cite{Minamitsuji:2010fp}
\bibitem{Minamitsuji:2010fp}
  M.~Minamitsuji, N.~Ohta and K.~Uzawa,
  ``Dynamical solutions in the 3-Form Field Background in the 
  Nishino-Salam-Sezgin Model,''
  Phys.\ Rev.\ D {\bf 81} (2010) 126005
  [arXiv:1003.5967 [hep-th]].

%\cite{Maeda:2010aj}
\bibitem{Maeda:2010aj}
  K.~-i.~Maeda, M.~Minamitsuji, N.~Ohta and K.~Uzawa,
  ``Dynamical $p$-branes with a cosmological constant,''
  Phys.\ Rev.\ D {\bf 82} (2010) 046007
  [arXiv:1006.2306 [hep-th]].
  
%\cite{Minamitsuji:2010kb}
\bibitem{Minamitsuji:2010kb}
  M.~Minamitsuji, N.~Ohta and K.~Uzawa,
  ``Cosmological intersecting brane solutions,''
  Phys.\ Rev.\ D {\bf 82} (2010) 086002
  [arXiv:1007.1762 [hep-th]].

%\cite{Nozawa:2010zg}
\bibitem{Nozawa:2010zg}
  M.~Nozawa and K.~i.~Maeda,
  ``Cosmological rotating black holes in five-dimensional fake supergravity,''
  Phys.\ Rev.\  D {\bf 83} (2011) 024018
  [arXiv:1009.3688 [hep-th]].

%\cite{Minamitsuji:2010uz}
\bibitem{Minamitsuji:2010uz}
  M.~Minamitsuji and K.~Uzawa,
  ``Cosmology in $p$-brane systems,''
  Phys.\ Rev.\ D {\bf 83} (2011) 086002
  [arXiv:1011.2376 [hep-th]].

%\cite{Maeda:2011sh}
\bibitem{Maeda:2011sh}
  K.~-i.~Maeda and M.~Nozawa,
  ``Black hole solutions in string theory,''
  Prog.\ Theor.\ Phys.\ Suppl.\  {\bf 189} (2011) 310
  [arXiv:1104.1849 [hep-th]].
  
%\cite{Minamitsuji:2011jt}
\bibitem{Minamitsuji:2011jt}
  M.~Minamitsuji and K.~Uzawa,
  ``Dynamics of partially localized brane systems,''
  Phys.\ Rev.\ D {\bf 84} (2011) 126006
  [arXiv:1109.1415 [hep-th]].
  
%\cite{Maeda:2012xb}
\bibitem{Maeda:2012xb}
  K.~-i.~Maeda and K.~Uzawa,
  ``Dynamical brane with angles: Collision of the universes,''
  Phys.\ Rev.\ D {\bf 85} (2012) 086004
  [arXiv:1201.3213 [hep-th]].

%\cite{Minamitsuji:2012if}
\bibitem{Minamitsuji:2012if}
  M.~Minamitsuji and K.~Uzawa,
  ``Cosmological brane systems in warped spacetime,''
  Phys.\ Rev.\ D {\bf 87} (2013) 046010
  [arXiv:1207.4334 [hep-th]].
  
\bibitem{LP}
R.~M.~Hornreich, M.~Luban and S.~Shtrikman, 
``Critical behavior at the onset of $\vec{k}$-space instability on the $\lambda$ line,'' 
Phys.\ Rev.\ Lett.\ 35 (1975) 1678.   
  
\bibitem{Kachru}
  S.~Kachru, X.~Liu and M.~Mulligan,
  ``Gravity Duals of Lifshitz-like Fixed Points,''  
Phys.\ Rev.\ D {\bf 78} (2008) 106005 [arXiv:0808.1725 [hep-th]].   

\bibitem{BN}
   K.~Balasubramanian and K.~Narayan,
   ``Lifshitz spacetimes from AdS null and cosmological solutions,''  
 JHEP {\bf 1008} (2010) 014 [arXiv:1005.3291 [hep-th]].  

\bibitem{DG}
  A.~Donos and J.~P.~Gauntlett,
  ``Lifshitz Solutions of D=10 and D=11 supergravity,''  
JHEP {\bf 1012} (2010) 002 [arXiv:1008.2062 [hep-th]].    

\bibitem{CH}
  W.~Chemissany and J.~Hartong,
  ``From D3-Branes to Lifshitz Space-Times,''
  Class.\ Quant.\ Grav.\  {\bf 28} (2011) 195011 [arXiv:1105.0612 [hep-th]].

%\cite{Uzawa:2013koa}
\bibitem{Uzawa:2013koa}
  K.~Uzawa and K.~Yoshida,
  ``Dynamical Lifshitz-type solutions and aging phenomena,''
  Phys.\ Rev.\ D {\bf 87} (2013) 106003
  [arXiv:1302.5224 [hep-th]].

%\cite{Youm:1999ti}
\bibitem{Youm:1999ti}
  D.~Youm,
  ``Partially localized intersecting BPS branes,''
  Nucl.\ Phys.\ B {\bf 556} (1999) 222 [hep-th/9902208].

%\cite{Cvetic:2000cj}
\bibitem{Cvetic:2000cj}
  M.~Cvetic, H.~Lu, C.~N.~Pope and J.~F.~Vazquez-Poritz, 
  ``AdS in warped space-times,''
  Phys.\ Rev.\ D {\bf 62} (2000) 122003
  [hep-th/0005246].
  
%\cite{Bergshoeff:1994cb}
\bibitem{Bergshoeff:1994cb}
  E.~Bergshoeff, R.~Kallosh and T.~Ortin,
  ``Duality versus supersymmetry and compactification,''  
  Phys.\ Rev.\ D {\bf 51} (1995) 3009  [hep-th/9410230].
  
%\cite{Bergshoeff:1995as}
\bibitem{Bergshoeff:1995as}
  E.~Bergshoeff, C.~M.~Hull and T.~Ortin,
  ``Duality in the type II superstring effective action,''
  Nucl.\ Phys.\  B {\bf 451} (1995) 547
  [arXiv:hep-th/9504081].

%\cite{Breckenridge:1996tt}
\bibitem{Breckenridge:1996tt}
  J.~C.~Breckenridge, G.~Michaud and R.~C.~Myers,
  ``More D-brane bound states,''
  Phys.\ Rev.\  D {\bf 55} (1997) 6438
  [arXiv:hep-th/9611174].

%\cite{Costa:1996zd}
\bibitem{Costa:1996zd}
  M.~S.~Costa and G.~Papadopoulos,
  ``Superstring dualities and $p$-brane bound states,''
  Nucl.\ Phys.\  B {\bf 510} (1998) 217
  [arXiv:hep-th/9612204].

%\cite{Burgess:2003mk}
\bibitem{Burgess:2003mk}
  C.~P.~Burgess, C.~Nunez, F.~Quevedo, G.~Tasinato and I.~Zavala,
  ``General brane geometries from scalar potentials: Gauged supergravities and accelerating universes,''
  JHEP {\bf 0308} (2003) 056
  [hep-th/0305211].
  
%\cite{Tolley:2006ht}
\bibitem{Tolley:2006ht}
  A.~J.~Tolley, C.~P.~Burgess, C.~de Rham and D.~Hoover,
  ``Scaling solutions to 6D gauged chiral supergravity,''
  New J.\ Phys.\  {\bf 8} (2006) 324
  [hep-th/0608083].

%\cite{Anchordoqui:2007sb}
\bibitem{Anchordoqui:2007sb}
  L.~Anchordoqui, H.~Goldberg, S.~Nawata and C.~Nunez,
  ``Cosmology from String Theory,''
  Phys.\ Rev.\ D {\bf 76} (2007) 126005
  [arXiv:0704.0928 [hep-ph]].
  
%\cite{Tolley:2007et}
\bibitem{Tolley:2007et}
  A.~J.~Tolley, C.~P.~Burgess, C.~de Rham and D.~Hoover,
  ``Exact Wave Solutions to 6D Gauged Chiral Supergravity,''
  JHEP {\bf 0807} (2008) 075
  [arXiv:0710.3769 [hep-th]].
  
\end{thebibliography}
\end{document}